\documentclass[aps,prd,12pt,floatfix,tightenlines,groupedaddress,nobalancelastpage,nofootinbib,amsmath]{revtex4}
\usepackage{epsfig}

\begin{document}

\title{Analytical treatment of neutrino asymmetry equilibration
from \\ flavour oscillations in the early universe}

\author{Yvonne Y. Y. Wong}\email{ywong@physics.udel.edu}\affiliation{
Department of Physics and Astronomy, University of Delaware,
Newark, Delaware 19716}


\begin{abstract}
A recent numerical study by A.~D.~Dolgov, S.~H.~Hansen, S.~Pastor,
S.~T.~Petcov, G.~G.~Raffelt, and D.~V.~Semikoz (DHPPRS) [Nucl.\
Phys.\ B (in press), hep-ph/0201287] found that complete or
partial equilibrium between all active neutrino flavours can be
achieved before the big bang nucleosynthesis epoch via flavour
oscillations, if the oscillation parameters are those inferred
from the atmospheric and solar neutrino data, and, in some cases,
if $\theta_{13}$ is also sizeable.  As such, cosmological
constraints on the electron neutrino-antineutrino asymmetry are
now applicable in all three neutrino sectors. In the present work,
we provide an analytical treatment of the scenarios considered in
DHPPRS, and demonstrate that their results are stable even for
very large initial asymmetries. The equilibration mechanism can be
understood in terms of an MSW-like effect for a maximally mixed
and effectively monochromatic system. We also comment on DHPPRS's
choices of mixing parameters, and their handling of collisional
effects, both of which could impinge on the extent of flavour
equilibrium.
\end{abstract}


 \maketitle

\section{Introduction}

One of the open questions in cosmology is the possibility of
admitting a large relic neutrino-antineutrino asymmetry. For an
ensemble of  neutrinos and antineutrinos of flavour $\alpha$ in
thermal and chemical equilibrium, this asymmetry may be
alternatively expressed in terms of the chemical potential
$\xi_{\nu_{\alpha}}$ of the species. As of now, there are no
direct observations of the cosmic neutrino background. Thus the
existence or otherwise of any sizeable $\xi$ can only be
established indirectly from requiring consistency with big bang
nucleosynthesis (BBN), and from the study of the cosmic microwave
background radiation (CMBR) spectrum.  Presently, BBN limits the
chemical potential in the $\nu_e$ sector to be at most of  order
$1$, while bounds on  $\xi_{\nu_{\mu}}$ and $\xi_{\nu_{\tau}}$ are
considerably less stringent \cite{bib:bbnbound}. A recent combined
analysis of BBN and CMBR has generated the constraints,
 $-0.01 < \xi_{\nu_e} <0.22$,
$|\xi_{\nu_{\mu}, \nu_{\tau}}| <2.6$, assuming no neutrino
oscillations  \cite{bib:cmbrbound}.

The situation changes if one considers also the effects of
neutrino oscillations.   In a new numerical study by A.~D.~Dolgov,
S.~H.~Hansen, S.~Pastor, S.~T.~Petcov, G.~G.~Raffelt, and
D.~V.~Semikoz (DHPPRS) \cite{bib:dhpprs}, it was shown that, for
the oscillation parameters inferred from the atmospheric neutrino
data and the large mixing angle (LMA) solution of the solar
neutrino problem, complete equilibrium between all active flavours
is established prior to the onset of BBN  at temperature $T \simeq
1 \ {\rm MeV}$.\footnote{Equilibration of the neutrino flavours
for the atmospheric and LMA oscillation parameters was first
suggested in Ref.\ \cite{bib:lunardini}.} For other solar neutrino
solutions, DHPPRS found that a partial equilibrium is possible if
the mixing angle $\theta_{13}$ is close to its present
experimental limit of $\tan^2 \theta_{13} \alt 0.065$. Clearly, if
flavour equilibrium holds, constraints on $\xi_{\nu_e}$ will apply
to all three flavours.

Central to the DHPPRS study is the inclusion of a highly nonlinear
neutrino self interaction potential in the evolution equation for
the ensemble. Previously, this term was found to give rise to
synchronised vacuum oscillations characterised strongly by the
initial conditions \cite{bib:kostelecky}. This raises some
questions: How sensitive are the DHPPRS results to the initial
asymmetries? For instance, can the presence of a large asymmetry
in the $\nu_{\mu}$ or $\nu_{\tau}$ sector delay its equilibration
with its $\nu_e$ counterpart?  Why do the final asymmetries
exhibit oscillatory behaviour in some but not all cases?
Asymmetries that oscillate out of phase with each other are
certainly not in equilibrium.

To find the answers, we begin with an analysis of two-flavour
oscillations involving $\nu_e$ in the solar neutrino parameter
space. The knowledge gained therefrom will be applied to the
three-flavour case, and, in particular, to investigating the role
of $\theta_{13}$. Whenever numerical quantities are called for,
e.g., the mass squared differences, we shall adopt the values used
in DHPPRS. These will be noted at the appropriate points. With the
exception of the small mixing angle (SMA) solution, the solar
mixing angle is always taken to be maximal in DHPPRS. This
provides  a motivation for us to examine also the more realistic
case of large but not maximal mixing.

\section{Two flavours}
\label{sec:twoflavours}

\subsection{Preliminary considerations}

Consider a two-state system consisting of $\nu_e$ and $\nu_x$,
where $\nu_x$ may be $\nu_\mu$, $\nu_\tau$, or a linear
combination thereof. We parameterise the transformation between
the weak and mass eigenstates in vacuum with a mixing angle
$\theta$,
\begin{equation}
\left(\begin{array}{c}
        \nu_e \\ \nu_x \end{array} \right)
= \left( \begin{array}{cc}
            \cos \theta & \sin \theta \\
            -\sin \theta & \cos \theta \end{array} \right)
\left(\begin{array}{c}
            \nu_1 \\ \nu_2 \end{array} \right),
\end{equation}
where the states $\nu_{1,2}$ have masses $m_{1,2}$ respectively.
The same parameterisation applies to the $\bar{\nu}_e
\leftrightarrow \bar{\nu}_x$ system.

For each momentum $p$, we write down the one-body reduced density
matrices $\rho$ ($\bar{\rho}$ for antineutrinos) and express them
in terms of the function $P_0$ ($\bar{P}_0$) and a
``polarisation'' vector ${\bf P}$ ($\bar{\bf P}$):
\begin{eqnarray}
\rho & =&  \left( \begin{array}{cc}
                \rho_{ee} & \rho_{ex} \\
                \rho_{xe} & \rho_{xx} \end{array} \right) =
                \frac{1}{2} \left[ P_0 + {\bf P} \cdot
                {\bf \sigma} \right], \nonumber \\
\bar\rho & =&  \left( \begin{array}{cc}
                \bar\rho_{ee} & \bar\rho_{xe} \\
                \bar\rho_{ex} & \bar\rho_{xx} \end{array} \right) =
                \frac{1}{2} \left[ \bar{P}_0 + \bar{\bf P} \cdot {\bf
                \sigma} \right],
\end{eqnarray}
where ${\bf P}= P_x {\bf x} + P_y {\bf y} + P_z {\bf z}$, and
${\bf \sigma} = \sigma_x {\bf x} + \sigma_y {\bf y} + \sigma_z
{\bf z}$ are the Pauli matrices.  In this notation, the $\nu_e$
and $\nu_x$ distribution functions at $p$ are respectively
\begin{eqnarray}
\label{eq:distributions} f_{\nu_e} &=& \frac{1}{2} \left[P_0 + P_z
\right] f_{\rm eq}(0), \nonumber \\ f_{\nu_x} &=& \frac{1}{2}
\left[P_0 - P_z \right] f_{\rm eq} (0),
\end{eqnarray}
for which we have chosen the reference distribution function
$f_{\rm eq}(0)$ to be of Fermi-Dirac form,
\begin{equation}
f_{\rm eq}(\xi) \equiv \frac{1}{1+ e^{p/T - \xi}},
\end{equation}
with chemical potential $\xi$ set to zero for all temperatures
$T$.  The distributions $f_{\bar{\nu}_e}$ and $f_{\bar{\nu}_x}$
may be similarly established from Eq.\ (\ref{eq:distributions}) by
replacing $P_0$ and $P_z$ with their antineutrino counterparts.
The number density of a particle species $\psi$ follows from
taking the integral of $f_{\psi}$ over all momenta, $n_{\psi}=
(1/2 \pi^2) \int f_{\psi} p^2 dp$. Note that the functions $(P_0,
{\bf P})$ and $(\bar{P}_0, \bar{\bf P})$ carry both momentum and
time dependence unless otherwise stated.

The evolution of ${\bf P}$ and $\bar{\bf P}$ is governed by the
quantum kinetic equations (QKEs) \cite{bib:sigl}
\begin{eqnarray}
\label{eq:dpdt}
 \dot{\bf P}\! \! \! & =  & \! \! \! +  \! \! \left[ \frac{\Delta m^2}{2p}
{\bf B} \! -\! \frac{8 \sqrt{2} G_F p}{3 m^2_W} E_{ee} {\bf z}
\right] \! \! \times \!{\bf P} + \sqrt{2} G_F ({\bf J} \! - \!
\bar{\bf J})\! \! \times \! {\bf P}, \nonumber
\\ \dot{\bar{\bf P}} \! \! \! & =  & \! \! \! - \! \! \left[ \frac{\Delta m^2}{2p}
{\bf B} \! - \! \frac{8 \sqrt{2} G_F p}{3 m^2_W} E_{ee} {\bf z}
\right] \! \! \times \!  \bar{\bf P} +  \sqrt{2} G_F({\bf J } \! -
\!  \bar{\bf J}) \! \! \times \!
 \bar{\bf P},\nonumber \\
\end{eqnarray}
where $\Delta m^2=m^2_2-m^2_1$, ${\bf B}=\sin 2 \theta \ {\bf
x}-\cos 2 \theta \ {\bf z}$, $G_F$ is the Fermi constant, $m_W$ is
the mass of the $W$ boson, $E_{ee}=(7/60)\pi^2T^4$ is the
electron-positron energy density, and we are assuming, at this
stage, that the background medium does not distinguish between
$\nu_{\mu}$ and $\nu_{\tau}$. Observe that refractive matter
effects for this system are entirely $CP$ symmetric --- we have
dropped the $CP$ asymmetric term proportional to the difference
between the charged lepton and antilepton number densities.  This
difference is expected to be of the order of the baryon asymmetry,
and is negligible in comparison with the $CP$ symmetric
background. The last term in Eq.\ (\ref{eq:dpdt}) comprising the
integrated polarisation vectors
 \begin{eqnarray}
 \label{eq:j}
{\bf J} &=& \frac{1}{2 \pi^2} \int  {\bf P} \ f_{\rm eq}(0) \ p^2
dp ,\nonumber \\ \bar{\bf J} &=& \frac{1}{2 \pi^2} \int \bar{\bf
P} \ f_{\rm eq}(0) \ p^2 dp,
\end{eqnarray}
arises from neutrino self interaction \cite{bib:pantaleone}, and
is by definition also time-dependent.

We have chosen to ignore the effects of collisions for now, since,
for the bulk of the momentum distribution, the neutrino mean free
path $\Gamma^{-1} \sim (G_F^2 p\ T^4)^{-1}$ is larger than an
effective oscillation length $\sim 2 \pi (V_x^2+V_z^2)^{-1/2}$,
where
\begin{equation}
\label{eq:vxvz}
V_x = \frac{\Delta m^2}{2p} \sin 2 \theta, \quad
V_z = - \frac{\Delta m^2}{2p} \cos 2 \theta - \frac{8 \sqrt{2} G_F
p}{3 m^2_W} E_{ee},
\end{equation}
given the mass squared differences and mixing angles of the
various solar neutrino solutions \cite{bib:bestfits}.
 The evolution of this two-state
system is therefore primarily oscillation-driven.   We also do not
consider repopulation from the background for consistency with the
$\Gamma=0$ assumption,  and hence $\dot{P}_0=\dot{\bar{P}}_0=0$.

Note that the collisionless approximation strictly  does not apply
to $\nu_{\mu} \leftrightarrow \nu_{\tau}$ oscillations in the
atmospheric parameter space. These oscillations become operational
at higher temperatures, where frequent ``non-forward'' scattering
on the background medium  modifies significantly the  system's
dynamics (i.e., the system is ``collision-driven''). One should
also be cautious when dealing with oscillation-driven systems
(such as $\nu_e \leftrightarrow \nu_x$) --- residual effects from
collisions may still be sizeable, especially if there is a
substantial period of large mixing.  This issue will be revisited
later in Sec.\ \ref{sec:large}, but let us emphasise at this point
that a full treatment of collisional effects on active-active
neutrino oscillations is considerably more complicated than is
implied in DHPPRS. The fact that both neutrino flavours can now
participate in momentum-changing scattering processes gives rise
to new terms in the evolution equation in addition to the simple
damping of $P_x$ and $P_y$ encountered in the active-sterile case.
See, for example, Ref.\ \cite{bib:mck&t} for details.

Ideally, we would like to compare the distribution functions of
the two neutrino species concerned.  When thermal and chemical
equilibria prevail, these are completely specified by the chemical
potentials $\xi_{\nu_e}=-\xi_{\bar{\nu}_e}$, and
$\xi_{\nu_x}=-\xi_{\bar{\nu}_x}$, which are related to the
neutrino-antineutrino asymmetries
$L_{\nu_{\alpha}}=(n_{\nu_{\alpha}}-n_{\bar{\nu}_{\alpha}})/n_{\gamma}$
via
\begin{equation}
\label{eq:asymmetry} L_{\nu_{\alpha}}  =   \frac{1}{12 \zeta(3)}
\left(\pi^2 \xi_{\nu_{\alpha}} + \xi^3_{\nu_{\alpha}} \right),
\end{equation} where $\alpha=e,x$, $n_{\gamma}=2
\zeta(3)T^3/\pi^2$ is the photon number density, and $\zeta$ is
the Riemann zeta function. A non-thermal distribution (e.g., due
to oscillations) generally has no well-defined chemical potential.
However, if we demand only that equality between the number
densities of $\nu_e$ and $\nu_x$ (and separately, $\bar{\nu}_e$
and $\bar{\nu}_x$) be established for flavour equilibration, then
it is sufficient to track  the evolution of the difference between
$L_{\nu_e}$ and $L_{\nu_x}$,
\begin{eqnarray}
\label{eq:le-lx} L_{\nu_e} - L_{\nu_x}  &=& \frac{1}{ 2 \pi^2
n_{\gamma}} \int \left[(f_{\nu_e} - f_{\bar{\nu}_e}) -
(f_{\nu_{x}} - f_{\bar{\nu}_{x}}) \right] p^2 dp \nonumber \\ &=&
\frac{1}{2 \pi^2 n_{\gamma}} \int (P_z - \bar{P}_z ) f_{\rm eq}(0)
p^2 dp \nonumber \\ &=& \frac{1}{n_{\gamma}} (J_z-\bar{J}_z),
\end{eqnarray}
regardless of whether thermal and/or chemical equilibria are in
place.

\subsection{Derivation of a collective evolution equation}
\label{sec:derivation}

Let us now examine the time development of the ``normalised''
vector ${\bf I} \equiv  ({\bf J} - \bar{\bf J})/n_{\gamma}$, for
which we construct from Eqs.\ (\ref{eq:dpdt}) and (\ref{eq:j}) an
evolution equation
\begin{eqnarray}
\label{eq:didtexact} \dot{{\bf I}}  &=&   {\bf B} \times
\frac{1}{2 \pi^2 n_{\gamma}} \int  \frac{\Delta m^2}{2p} ({\bf P}
+ \bar{\bf P})  f_{\rm eq}(0) p^2 dp \nonumber \\ && \hspace{3mm}
- {\bf z} \times \frac{1}{2 \pi^2 n_{\gamma}} \int \frac{8
\sqrt{2} G_F p}{3 m^2_W} E_{ee} ( {\bf P} + \bar{\bf P} ) f_{\rm
eq}(0) p^2 dp.\nonumber \\
\end{eqnarray}
Equilibration between the two neutrino species will necessarily
imply ${I}_z=0$.
 Note, however, that the converse is not
generally true.

Equation (\ref{eq:didtexact}) is exact;  our task is to find
approximate solutions for ${\bf P}$ and $\bar{\bf P}$. To this
end, we first rewrite the evolution equation (\ref{eq:dpdt}) in
matrix form,
\begin{widetext}
\begin{equation}
\label{eq:dpdtmatrix} \frac{d}{dt} \left( \begin{array}{c}
                        P_x \\ P_y \\ P_z \end{array} \right)
= \left[ \left( \begin{array}{ccc}
            0 & -V_z & 0 \\
            V_z & 0 & - V_x \\
            0 & V_x & 0 \end{array} \right)
+  \sqrt{2} G_F n_{\gamma} \left( \begin{array}{ccc}
            0 & -{I}_z & {I}_y \\
            {I}_z & 0 & - {I}_x \\
            -{I}_y & {I}_x & 0 \end{array} \right) \right]
  \left( \begin{array}{c}
            P_x \\ P_y \\ P_z \end{array} \right)
  = ( {\cal V} + {\cal S} ) {\bf P}.
\end{equation}
\end{widetext}
where $V_x$ and $V_z$ are as defined in Eq.\ (\ref{eq:vxvz}). For
the antineutrino system, the matrix ${\cal V}$ is replaced with
$-{\cal V}$.

Before proceeding, observe that in the absence of the self
interaction term ${\cal S}$, the vector ${\bf P}$ would simply
exhibit the usual ``matter-suppressed'' precession around the
vector ${\bf V}=V_x {\bf x}+V_z {\bf z}$ at a rate
\begin{equation}
\omega_{\cal V}=\sqrt{V_x^2 + V_z^2},
\end{equation}
and gradually give way to vacuum oscillations as the
electron-positron energy density drops off with the expansion of
the universe. If, hypothetically, only the ${\cal S}$ term is
present, ${\bf P}$ would precess around ${\bf I}$ at a rate
\begin{equation}
\omega_{\cal S}=\sqrt{2} G_F  n_{\gamma}|{\bf I}|,
\end{equation}
provided that ${\bf I}$ changes sufficiently slowly with time (to
be justified later).

In the following, we shall work under the assumption that the
inequality
\begin{equation}
\label{eq:synchcondition}
\omega_{\cal V} \ll \omega_{\cal S},
\end{equation}
is maintained for all momenta at all times. This condition
translates roughly into requiring that
\begin{equation}
\label{eq:t>tr}
 1  \gg  \frac{8 p}{3 m^2_W n_{\gamma}  |{\bf I}|}
\simeq 2 \times 10^{-9} \ \frac{y}{|{\bf I}|}
 \left(\frac{T}{\rm MeV} \right)^2,
\end{equation}
for $T>T_{\rm r}$, and
\begin{equation}
\label{eq:t<tr}
1 \gg  \frac{\Delta m^2}{2 \sqrt{2} G_F p \
n_{\gamma} |{\bf I}|}
 \simeq 0.12 \, \, \frac{y^{-1}}{|{\bf I}|} \left(
\frac{|\Delta m^2|}{{\rm eV}^2} \right)\left( \frac{T}{\rm MeV}
\right)^{-4},
\end{equation}
for $T<T_{\rm r}$, where $y \equiv p/T$, and
\begin{equation}
\label{eq:tref} T_{\rm r} \simeq 19.8\  \, y^{-\frac{1}{3}}\left(
\frac{|\Delta m^2|}{{\rm eV}^2} \right)^{\frac{1}{6}} \, {\rm
MeV},
\end{equation}
is the temperature at which the vacuum and electron-positron
background terms become equal in magnitude.  The quantity $|{\bf
I}|$ may be regarded crudely as a measurement of the degree of
alignment of the individual polarisation vectors ${\bf P}$ and
$\bar{\bf P}$, such that if all neutrinos and antineutrinos are in
flavour eigenstates (i.e., $P_{x,y}\simeq\bar{P}_{x,y}\simeq0$),
the magnitude of ${\bf I}$ equals the difference in the $\nu_e$
and $\nu_x$ neutrino-antineutrino asymmetries.  It turns out that
initial fulfilment of the requirement (\ref{eq:synchcondition})
tends to preserve the alignment until the self interaction term
significantly weakens with the expansion of the universe through
$n_{\gamma}$. Hence, assuming all ${\bf P}^{\rm i}$ and $\bar{\bf
P}^{\rm i}$, where the superscript ``i'' denotes initial, to be in
the ${\bf z}$ direction, the reader may substitute in Eqs.\
(\ref{eq:t>tr}) and (\ref{eq:t<tr}) the relation $|{\bf
I}|\simeq|{\bf I}^{\rm i}|\simeq|L^{\rm i}_{\nu_e}-L^{\rm
i}_{\nu_x}|$, and  see that these conditions are always met by the
bulk of the momentum distribution for the $\Delta m^2$'s
concerned, provided that the disparity between the initial
asymmetries is, say, $\gtrsim10^{-5}$ in magnitude.\footnote{The
case of equal initial asymmetries has {\it a priori} no bearing
for the purpose of the present work, although it could lead to
some very interesting phenomena.}

We solve the three coupled differential equations
(\ref{eq:dpdtmatrix}) by transforming to an instantaneous diagonal
basis defined as ${\bf Q}={\cal U}{\bf P}$, in which the evolution
equation takes the form
\begin{equation}
\label{eq:dqdt} \dot{\bf Q} = \left[ {\cal U} ({\cal V} + {\cal
S}) {\cal U}^{-1} - {\cal U} \dot{\cal U}^{-1} \right] {\bf Q},
\end{equation}
with ${\cal U}({\cal V}+{\cal S}){\cal U}^{-1}={\rm Diag}(k_1,\
k_2,\ k_3)$, where the eigenvalue $k_{1}$ is identically zero, and
$k_{2,3}$ are two imaginary numbers of equal magnitude but
opposite signs. The associating eigenvectors ${\bf q}_{1,2,3}$ in
terms of the original coordinates $\{{\bf x},{\bf y},{\bf z} \}$
constitute the columns of the transformation matrix
\begin{equation}
{\cal U}^{-1} = {\cal U}^{\dagger} =\left(\begin{array}{c|c|c}
 {\bf q}_1 & {\bf q}_2 & {\bf q}_3 \end{array} \right),
\end{equation}
where ${\bf q}_1$ is real, and ${\bf q}_{2,3}$  form a complex
conjugate (c.c.) pair.  In the limit $\omega_{\cal
V}\ll\omega_{\cal S}$, the eigenvalues are
\begin{equation}
k_1 =0, \qquad k_2 = k_3^* \simeq i \omega_{\cal S}  + {\cal
O}(\omega_{\cal V}),
\end{equation}
and the real eigenvector ${\bf q}_1$ is well approximated by
\begin{eqnarray}
\label{eq:q1} {\bf q}_1 &\simeq& \frac{1}{\sqrt{I_x^2+I_y^2+
I_z^2}} \left(
\begin{array}{c}
    I_x \\ I_y \\ I_z \end{array} \right) + \left[{\cal O}(\omega_{\cal V}/\omega_{\cal
    S}) {\bf q}_2^0 + {\rm c.c.} \right] \nonumber \\ &\equiv& \hat{\bf I} +
    \left[ {\cal O}(\omega_{\cal V}/\omega_{\cal
    S}) {\bf q}_2^0 + {\rm c.c.} \right],
\end{eqnarray}
where $\{{\bf q}_1^0, {\bf q}^0_2, {\bf q}^0_3 \}$ are the set of
eigenvectors in the limit $\omega_{\cal V}=0$.  We shall not
reproduce here the exact forms of the remaining two complex
conjugate eigenvectors, but simply point out that ${\bf q}_2$ and
${\bf q}_3$ together sweep out a plane perpendicular to ${\bf
q}_1$.

Equation (\ref{eq:dqdt}) is not yet soluble;  the term  ${\cal
U}\dot{\cal U}^{-1}$ contains off-diagonal elements.  However,
these may be set to zero to facilitate calculations since the
condition
\begin{equation}
\chi_{ij} \equiv \left| \frac{({\cal U} \dot{\cal
U}^{-1})_{ij}}{k_i-k_j} \right| \sim \frac{1}{\omega_{\cal S}}
\frac{1}{|{\bf I}|} \left| \frac{d {\bf I}}{dt} \right|  \ll 1 ,
\end{equation}
where $i,j=1,2,3$ and $i\neq j$, is always satisfied in the
$\omega_{\cal V}\ll\omega_{\cal S}$ limit.  This can be seen from
Eq.\ ({\ref{eq:didtexact}), in which any one component of
$\dot{\bf I}$ is at most of order $\omega_{\cal V}|{\bf I}|$, such
that $\chi_{ij}\sim\omega_{\cal V}/\omega_{\cal S}\ll1$ is
consistent with earlier assumptions.

The formal solution to the now decoupled evolution equation
(\ref{eq:dqdt}) is
\begin{equation}
{\bf Q} \simeq {\rm Diag}(1,\ e^{i \int^t_{t_{\rm i}} \omega_{\cal
S} dt'},\ e^{-i \int^t_{t_{\rm i}} \omega_{\cal S} dt'}) {\bf
Q}^{\rm i},
\end{equation}
or equivalently in the original $\{{\bf x, y, z} \}$ basis,
\begin{eqnarray}
\label{eq:psolution} {\bf P} &\simeq& {\cal U}^{-1} {\rm Diag}(1,\
e^{i\int^t_{t_{\rm i}} \omega_{\cal S} dt'},\ e^{-i \int^t_{t_{\rm
i}} \omega_{\cal S} dt'}) {\cal U}^{\rm i} {\bf P}^{\rm i}
\nonumber
\\ &=& \left( \begin{array}{c|c|c} {\bf q}_1 & {\bf q}_2 & {\bf q}_3  \end{array}
\right) \left(\begin{array}{c}
         {\bf q}_1^{\rm i} \cdot {\bf P}^{\rm i} \\
         e^{i \int^t_{t_{\rm i}} \omega_{\cal S} dt'}\
            {{\bf q}_2^{\rm i}}^* \cdot {\bf P}^{\rm i} \\
         e^{-i \int^t_{t_{\rm i}} \omega_{\cal S} dt'} \
            {{\bf q}_3^{\rm i}}^* \cdot {\bf P}^{\rm i} \end{array}
         \right),
\end{eqnarray}
where the superscript/subscript ``i'' denotes initial. Observe
that the terms $e^{i \int \omega_{\cal S} dt}$ and $e^{-i \int
\omega_{\cal S} dt}$ lead to rapid oscillations which average to
zero over the characteristic time-scale of ${\bf q}_{1,2,3}$. We
retain only the time-averaged component and adopt the identity
(\ref{eq:q1}), whereupon ${\bf P}$ becomes
\begin{equation}
\label{eq:averagedP} {\bf P} \simeq ( {\bf P}^{\rm i} \cdot
\hat{{\bf I}}^{\rm i} ) \hat{\bf I} + {\cal O}(\omega_{\cal
V}/\omega_{\cal S}) \hat{\bf I} + \left[ {\cal O}(\omega_{\cal
V}/\omega_{\cal S}) {\bf q}_2^0 + {\rm c.c.} \right],
\end{equation}
which, to the lowest order in $\omega_{\cal V}/\omega_{\cal S}$,
is a product of two quantities ${\bf P}^{\rm i} \cdot \hat{{\bf
I}}^{\rm i}$ and $\hat{\bf I}$ carrying respectively the momentum
and the time dependences of the original function.  Equation
(\ref{eq:averagedP})  pertains also to $\bar{\bf P}$, save for the
replacement of ${\bf P}^{\rm i}$ with $\bar{\bf P}^{\rm i}$, and
$\omega_{\cal V}$ with $-\omega_{\cal V}$. Substituting into Eq.\
(\ref{eq:didtexact}), we obtain for the collective vector ${\bf
I}$ a simplified evolution equation,
\begin{eqnarray}
\label{eq:didtapprox}
 \dot{{\bf I}} & \simeq & \frac{1}{|{\bf I}|}
\frac{1}{2 \pi^2 n_{\gamma}} \left[ {\bf B} \! \! \int  \!
\frac{\Delta m^2}{2 p} ({\bf P}^{\rm i} +\bar{\bf P}^{\rm i})
\cdot \hat{\bf I}^{\rm i} f_{\rm eq}(0) p^2 dp \right. \nonumber
\\ &&  \hspace{5mm} \left. - {\bf z} \! \! \int \!
\frac{8 \sqrt{2} G_F p}{3 m^2_W} E_{ee}
({\bf P}^{\rm i} +\bar{\bf P}^{\rm i}) \cdot \hat{\bf I}^{\rm i}
f_{\rm eq}(0) p^2 dp \right] \times {{\bf I}}, \nonumber
\\
\end{eqnarray}
to the lowest order in $\omega_{\cal V}/\omega_{\cal S}$.

The non-dissipative character of ${\bf I}$'s evolvement  is
immediately manifest in Eq.\ (\ref{eq:didtapprox}).  Furthermore,
the dynamics of the neutrino-antineutrino ensemble can be
completely and simply determined from the initial conditions and
from known external factors, independently of the evolution of the
ensemble {\it per se}.  Synchronised vacuum oscillations in
multi-momentum systems subject to intense self interactions was
discovered numerically in Ref.\ \cite{bib:kostelecky}, and
recently reinterpreted in Ref.\ \cite{bib:synch}.  Equation
(\ref{eq:didtapprox}) is essentially a generalisation of the
physical picture developed in Ref.\ \cite{bib:synch} for pure
vacuum oscillations. We opted to conduct a more systematic,
first-principles derivation here for pedagogy.

We take as the initial condition that all neutrinos and
antineutrinos are in flavour eigenstates such that ${\bf P}^{\rm
i}$ and $\bar{\bf P}^{\rm i}$ are parallel to $\hat{\bf I}^{\rm
i}$, and
\begin{equation}
{\bf P}^{\rm i} \cdot \hat{\bf I}^{\rm i} \simeq \frac{f^{\rm
i}_{\nu_e} - f^{\rm i}_{\nu_{x}}}{f_{\rm eq}(0)}, \qquad \bar{\bf
P}^{\rm i} \cdot \hat{\bf I}^{\rm i} \simeq \frac{f^{\rm
i}_{\bar{\nu}_e} - f^{\rm i}_{\bar{\nu}_{x}}}{f_{\rm eq}(0)},
\end{equation}
up to a common sign.  Two additional assumptions of thermal as
well as chemical equilibria, that is,
\begin{eqnarray}
f^{\rm i}_{\nu_e}  \simeq  f_{\rm eq}(\xi^{\rm i}_{\nu_e}), &
\qquad & f^{\rm i}_{\bar{\nu}_e} \simeq  f_{\rm eq}(-\xi^{\rm
i}_{\nu_e}), \nonumber \\ f^{\rm i}_{\nu_{x}}  \simeq  f_{\rm
eq}(\xi^{\rm i}_{\nu_{x}}), & \qquad & f^{\rm i}_{\bar{\nu}_{x}}
\simeq f_{\rm eq}(-\xi^{\rm i}_{\nu_{x}}),
\end{eqnarray}
then allow for a further approximation of Eq.\
(\ref{eq:didtapprox}) as
\begin{widetext}
\begin{eqnarray}
\label{eq:didtlong} \lefteqn{\dot{{\bf I}}} &\simeq&
\frac{1}{|{\bf I}|} \frac{1}{2 \pi^2 n_{\gamma}} \left\{ {\bf B}
\! \int \! \! \frac{\Delta m^2}{2 p} \left[f_{\rm eq}(\xi^{\rm
i}_{\nu_e}) + f_{\rm eq}(-\xi^{\rm i}_{\nu_e}) - f_{\rm
eq}(\xi^{\rm i}_{\nu_{x}}) - f_{\rm eq}(-\xi^{\rm i}_{\nu_{x}})
\right] p^2 dp \right. \nonumber
\\ && \hspace{10mm} \left. - {\bf z} \! \int \!  \frac{8 \sqrt{2} G_F p}{3
m^2_W} E_{ee} \left[f_{\rm eq}(\xi^{\rm i}_{\nu_e}) + f_{\rm
eq}(-\xi^{\rm i}_{\nu_e}) - f_{\rm eq}(\xi^{\rm i}_{\nu_{x}}) -
f_{\rm eq}(-\xi^{\rm i}_{\nu_{x}}) \right] p^2 dp \right\} \!
\times {\bf I}.
\end{eqnarray}
The integrals are now in a form opportune for the exploitation of
these very useful identities:
\begin{eqnarray}
\int [ f_{\rm eq}(\xi) - f_{\rm eq} (-\xi)] p^2 dp  & = &
\frac{T^3}{3} ( \pi^2 \xi + \xi^3), \nonumber \\  \int  [ f_{\rm
eq}(\xi) + f_{\rm eq} (-\xi)] p \ dp  &= & \frac{T^2}{6} ( \pi^2 +
3 \xi^2 ), \nonumber
\\ \int  [ f_{\rm eq}(\xi)  +  f_{\rm eq} (-\xi)] p^3  dp  &= &
\frac{T^4}{60} (7 \pi^4 +  30 \pi^2 \xi^2 + 15 \xi^4). \nonumber
\\
\end{eqnarray}
These, together with the conservation of the collective vector
${\bf I}$,
\begin{eqnarray}
|{\bf I}| \simeq |{\bf I}^{\rm i}| \! \! &\simeq& \! \! \frac{1}{2
\pi^2 n_{\gamma}} \left| \int [(f^{\rm i}_{\nu_e} - f^{\rm
i}_{\bar\nu_e}) - (f^{\rm i}_{\nu_{x}} - f^{\rm i}_{\bar\nu_{x}})]
p^2 dp \right| \nonumber
\\ \! \! &\simeq& \! \! \frac{T^3}{6 \pi^2 n_{\gamma}}
\left| \ \pi^2 (\xi^{\rm i}_{\nu_e} -\xi^{\rm i}_{\nu_{x}}) +
({\xi^{\rm i}_{\nu_e}}^3 - {\xi^{\rm i}_{\nu_{x}}}^3) \right|,
\end{eqnarray}
 permit us to recast Eq.\ (\ref{eq:didtlong})
into a more illuminating and readily soluble form,
\begin{equation}
\label{eq:didtfinal}
 \dot{{\bf I}} \simeq \frac{3}{2}\frac{\widetilde{y} \
  ({\xi^{\rm i}_{\nu_e}}^2 -
{\xi^{\rm i}_{\nu_{x}}}^2 )}{| \pi^2 (\xi^{\rm i}_{\nu_e}
-\xi^{\rm i}_{\nu_{x}}) +  ({\xi^{\rm i}_{\nu_e}}^3 - {\xi^{\rm
i}_{\nu_{x}}}^3) |}  \left( \frac{\Delta m^2}{2 \widetilde{p}}
{\bf B} - \frac{8 \sqrt{2} G_F \widetilde{p}}{3 m^2_W} E_{ee} {\bf
z} \right) \times {\bf I},
\end{equation}
\end{widetext}
where
\begin{equation}
\label{eq:tildep}
 \widetilde{y} = \frac{\widetilde{p}}{T}
\equiv \sqrt{\pi^2 + \frac{1}{2} \left( {\xi^{\rm i}_{\nu_e}}^2 +
{\xi^{\rm i}_{\nu_{x}}}^2 \right)},
\end{equation}
represents some ``average'' momentum.  Equations
(\ref{eq:didtfinal}) and (\ref{eq:tildep}) will form the basis of
the discussions to follow.\footnote{The method used to derive Eq.\
(\ref{eq:didtfinal}) from the QKEs is equally valid had a $CP$
asymmetric background been included. In fact, even
 if the background strongly differentiates between neutrinos and
antineutrinos, the resulting evolution equation would
 still predict identical behaviours for both $CP$ partners, contrary to
the case when self interaction is absent. Unfortunately, this
scenario is of little interest for the early universe. Perhaps it
might find application in a supernova environment.}

\subsection{Discussions}

Equation  (\ref{eq:didtfinal}) has a straightforward
interpretation.  Consider the terms inside the parentheses. These
are identically the vacuum and electron-positron background terms
found in a typical single momentum evolution equation [Eq.\
(\ref{eq:dpdt}) minus self interaction], and control the
ensemble's ``collective'' matter-affected mixing angle,
\begin{equation}
\sin 2 \theta_c = \left. \frac{V_x}{\sqrt{V_x^2 + V_z^2}}
\right|_{p= \widetilde{p}},
\end{equation}
where $V_x$ and $V_z$ are evaluated for $p= \widetilde{p}$.

The characteristic momentum $\widetilde{p}$ defined in Eq.\
(\ref{eq:tildep}) reflects, to some extent, the initial
configuration of the ensemble. In addition to augmenting the
neutrino-antineutrino asymmetry, a large positive chemical
potential for a neutrino flavour tends to skew the distribution
towards higher momenta, while leaving the spectrum of its less
abundant antineutrino virtually intact.  A large negative chemical
potential has the opposite effect.  Hence the net result for
$\widetilde{p}$ is that it follows more or less the trends of the
more abundant species, and thereby grows with $|\xi^{\rm i}|$. For
initial chemical potentials satisfying the constraint
\begin{equation}
\label{eq:constraint}
{\xi^{\rm i}_{\nu_e}}^2 + {\xi^{\rm
i}_{\nu_x}}^2 \alt 2 \pi^2,
\end{equation}
the value of $\widetilde{p} \simeq \pi T$ originates from the
mismatch between the averages $\langle p \rangle$ and $\langle 1/p
\rangle^{-1}$ taken over the function $d^2f_{\rm eq}(\xi)/d\xi^2$
evaluated at $\xi=0$.

The quantity exhibiting by far the strongest dependence on the
initial conditions is the collective matter-affected oscillation
length, or equivalently, the collective effective mass squared
difference,
\begin{equation}
\Delta m^2_{\rm eff} = \kappa \left. 2 \widetilde{p}\ \sqrt{V_x^2
+ V_z^2} \right|_{p = \widetilde{p}},
\end{equation}
where
\begin{equation}
\label{eq:kappa}
\kappa = \frac{3}{2}\frac{\widetilde{y} \
({\xi^{\rm i}_{\nu_e}}^2 - {\xi^{\rm i}_{\nu_{x}}}^2 )}{| \pi^2
(\xi^{\rm i}_{\nu_e} -\xi^{\rm i}_{\nu_{x}}) +  ({\xi^{\rm
i}_{\nu_e}}^3 - {\xi^{\rm i}_{\nu_{x}}}^3) |}.
\end{equation}
For instance, $\Delta m^2_{\rm eff}$ vanishes for
$\xi_{\nu_e}^{\rm i}=- \xi_{\nu_x}^{\rm i}$, and oscillations are
switched off completely, as pointed out in DHPPRS. (Note again
that the present formulation is invalid for identical initial
asymmetries.)

An approximate solution to Eq.\ (\ref{eq:didtfinal}) can be
established in the adiabatic limit by mapping ${\bf I}$ onto an
instantaneous diagonal basis (in this case, isomorphic to an
instantaneous mass basis) and setting the time derivative of the
transformation matrix to zero. The resulting expression for the
variable $I_z$ is
\begin{equation}
\label{eq:Iz} I_z \simeq \left( c 2 \theta_c c 2 \theta_c^{\rm i}
+ s 2 \theta_c s 2 \theta^{\rm i}_c \ \cos \! \! \int^t_{t_{\rm
i}} \! \! \frac{\Delta m^2_{\rm eff}}{2 \widetilde p} dt' \right)
I_z^{\rm i},
\end{equation}
where $c 2 \theta=\cos 2 \theta$, $s 2 \theta=\sin 2 \theta$,
subject to the validity of the adiabatic condition
\begin{equation}
\label{eq:adparameter} \gamma \equiv \left|\frac{V_z \dot{V}_x -
V_x \dot{V}_z}{\kappa (V_x^2 + V_z^2)^{\frac{3}{2}}}
\right|_{p=\widetilde{p}} \alt 1,
\end{equation}
at all times.  For maximal mixing and the matter density profile
at hand, $\gamma$ is a maximum at $T= T_{\rm r}$ [Eq.\
(\ref{eq:tref}) with $y = \widetilde{y}$], at which it takes on a
particularly simple form:
\begin{equation}
\label{eq:gamma45} \left. \gamma \right|_{\theta = 45^{\circ}}
\simeq 1.5 \times 10^{-5} \ \kappa^{-1} \left(\frac{|\Delta
m^2|}{{\rm eV}^2} \right)^{-\frac{1}{2}},
\end{equation}
where we have used the relation $dT/dt \simeq - 5.44 T^3/m_{\rm
pl}$, and $m_{\rm pl} \simeq 1.22 \times 10^{22} \ {\rm MeV}$ is the Planck
mass.

Equation (\ref{eq:Iz}) predicts for maximal mixing an MSW-like
effect, transforming $I_z$ from $I_z^{\rm i}$ to $0$ (plus  some
small amplitude oscillations) when vacuum oscillations overcome
refractive matter effects (i.e., when $c 2 \theta_c \simeq c
2\theta$). The temperature at which this transition takes place is
given roughly by $T_{\rm r}$ in Eq.\ (\ref{eq:tref}) with
$y=\widetilde{y}$. For initial chemical potentials satisfying the
constraint (\ref{eq:constraint}), this turns out to be $\simeq
2.6\ {\rm MeV}$ for $\Delta m^2 = 4.5 \times 10^{-5}\ {\rm eV}^2$
(LMA). Extremely large initial $|\xi|$'s can lower $T_{\rm r}$
somewhat, but the connection is weak. As an illustration, the
setting of $\xi_{\nu_e}^{\rm i}=0$ and $\xi_{\nu_x}^{\rm i}=10$
gives, for the same $\Delta m^2$, $T_{\rm r} \simeq 1.9\ {\rm
MeV}$.  In the cases of the LOW ($\Delta m^2 = 1 \times 10^{-7}\
{\rm eV}^2$) and the Vacuum ($\Delta m^2 = 8 \times 10^{-11}\ {\rm
eV}^2$) solutions, the temperatures are $\simeq0.9 \ {\rm MeV}$
and $\simeq 0.3\ {\rm MeV}$ respectively, for reasonable
$|\xi^{\rm i}|$'s. Evidently, only the LMA transition can take
place well ahead of BBN.

For the oscillation parameters of the SMA solution ($\Delta m^2 =
7 \times 10^{-6}\ {\rm eV}^2$, $\sin 2 \theta =0.05$), $I_z$
remains close to its initial value even after the ``transition''
at $T_{\rm r}\simeq 1.9\ {\rm MeV}$, since both the vacuum
oscillation and matter refraction terms are predominantly in the
negative ${\bf z}$ direction for $\Delta m^2
> 0$, and the usual MSW resonance condition
cannot be satisfied. This has an important implication: in the
absence of substantial collisional damping and other effects not
considered here, any equilibration of two neutrino species in a
purely two-flavour scenario is, strictly speaking, an accident of
maximal mixing.

A second deciding factor on the efficacy of flavour equilibration
is the adiabaticity of the transition from matter-suppressed to
vacuum oscillations. Unlike that encountered in, for instance,
solar neutrino analyses, the adiabaticity parameter $\gamma$ of
Eq.\ (\ref{eq:adparameter}) is strongly dependent on the initial
conditions. Take for concreteness the case of $\xi_{\nu_e}^{\rm
i}=0$.   The reader may verify that for the LMA solution, the
condition (\ref{eq:gamma45}) is always true if $|\xi_{\nu_x}^{\rm
i}| \gtrsim 0.01$, but can be badly violated at the transition
point for the LOW $\Delta m^2$ unless $|\xi_{\nu_x}^{\rm i}|
\gtrsim 0.1$.

In the case of maximal mixing,  violation of adiabaticity at the
transition point generally results in large amplitude
``post-transition'' oscillations about the equilibrium point at an
angular frequency roughly equal to $\Delta m^2_{\rm eff}/2
\widetilde{p}$. Naturally, this is quite a separate phenomenon
from true equilibration.   On the other hand, an adiabaticity
parameter that evaluates to infinity at all times (because, for
example, $\kappa\to0$) signifies that there is no transition at
all.  From the perspective of equilibrating two vastly different
asymmetries, the requirement of $|\xi^{\rm i}_{\nu_x}| \gtrsim
0.01$ (assuming $\xi^{\rm i}_{\nu_e} = 0$) in the LMA case for a
smooth transition is, by definition, not a major concern. We
therefore  dwell no further on this topic, but simply point out to
the interested reader that there is an existing body of works
devoted to non-adiabatic transitions in the solar interior
\cite{bib:nonad}. The almost exponential density profile of the
sun is of course very different from the matter density profile
considered here. This will alter the post-transition oscillation
frequency and amplitude somewhat, but not the essential physics of
non-adiabatic effects.

\subsection{Large but not maximal mixing}
\label{sec:large}

Previously, we saw that equilibration of $\nu_e$ and $\nu_x$ in
the absence of collisional effects is peculiar to maximal mixing.
In particular, if $\Delta m^2$ corresponds to that of the LMA
solution to the solar neutrino problem, equilibrium is
complete before BBN. However, the real LMA solution encompasses a
range of mixing parameters that are merely large, but not maximal
\cite{bib:bestfits}. If we had used instead $\sin 2 \theta \simeq
0.88$, Eq.\ (\ref{eq:Iz}) would predict for this scenario only a
$\sim 50\%$ reduction in $I_z^{\rm i}$, i.e., a partial
equilibrium.

The question now is: can collisions save the scene?  A full
treatment of active-active collisions may be complicated. But for
the purpose of exploring the possibilities, we may want to try
including in the evolution equation (\ref{eq:dpdt}) a simple
active-sterile style damping term $-D (P_x + P_y)$ [and $-D
(\bar{P}_x + \bar{P}_y)$ for the antineutrinos], where $D=(1/2)
\lambda G_F^2 p \ T^4$, and $\lambda$ is a positive number
reflecting the relative amplitudes of the available inelastic
scattering and elastic scattering processes that distinguish
between the two flavours.

Introduction of this``toy'' damping term has a simple consequence
for the collective evolution equation (\ref{eq:didtfinal}): there
is now an extra term of the form
\begin{equation}
-\widetilde{D} (I_x + I_y),
\end{equation}
with $\widetilde{D} =  (27 \zeta(3)/2\pi^2) \lambda G_F^2 T^5$ for
$|\xi^{\rm i}| \alt 1$. For $\Delta m^2_{\rm eff}/2 \widetilde{p}
\gtrsim \widetilde{D}$, the corresponding solution is similar to
Eq.\ (\ref{eq:Iz}), but with a twist:
\begin{equation}
\label{eq:dampedIz}
I_z \simeq e^{-\int^t_{t_{\rm i}} \Omega dt'}
\!\! \cos 2 \theta_c \cos 2 \theta_c^{\rm i}  \ I_z^{\rm i} + {\rm
damped \, oscillations},
\end{equation}
where $\Omega \simeq \sin^2 2 \theta_c \widetilde{D}$ may be
interpreted as  an ``instantaneous equilibration rate''
\cite{bib:yyy}. Clearly, $\Omega$ scales with the amount of
mixing, causes the exponential in Eq.\ (\ref{eq:dampedIz}) to
decrease with time, and thus contributes to equilibrating the two
flavours concerned.  This equilibrating power is especially useful
if a partial equilibrium has already been reached via an MSW-like
effect.  However, the extent of the equilibrium is now dependent
on the nature of the scattering processes. As a crude estimate,
the setting of $\lambda=1$ will give $\exp[\cdots] \simeq 0.16$
for $\sin 2 \theta \simeq 0.88$ at $T \simeq 1 \ {\rm MeV}$. Thus,
together with the MSW-like transition, $I_z$ may still be able to
reach a value of $\simeq 0.08 \ I^{\rm i}_z$ prior to BBN.

Before closing this section, let us stress again that it is not
clear at this stage if a proper treatment of active-active
collisions will indeed lead to an outcome similar to that afforded
to us by the simple active-sterile picture.  The latter can only
serve as a rough guide.

\section{Three flavours}

We shall not derive from first principles a three-flavour analogue
of Eq.\ (\ref{eq:didtfinal}).  Such an exercise is perhaps not
worthwhile since the collisionless limit is strictly not
applicable at higher temperatures. Furthermore, a complete set of
QKEs for three active flavours incorporating all necessary
collision terms has yet to be written down in a user-friendly
form, and until we know how to handle its better established
two-flavour counterpart, we shall not dwell on the fine details of
the three-flavour case.

However, granted that the role of the self interaction term is to
cause the ensemble to behave in an effectively monochromatic
manner, a qualitative, or even semi-quantitative,  picture is still
available if we suppose that, like the two-flavour case, the
matter-affected mixing structure of the three-flavour system is
determined by the single mode Hamiltonian (in flavour space),
\begin{equation}
\label{eq:3ham} \frac{1}{2p}\ U \! \left(\!
\begin{array}{ccc}
                            m^2_1 \!&0& 0\\
                            0 & \!m^2_2\! & 0 \\
                            0 & 0 & \!m^2_3 \end{array}\! \right)\!
                             U^{\dagger} - \frac{8
\sqrt{2} G_F p}{3 m^2_W} \! \left(\!
\begin{array}{ccc}
        E_{ee} \! \!+ \! \!E_{\mu \mu} \!& 0 & 0 \\
        0 &\! E_{\mu \mu}\!& 0 \\
        0 & 0 & 0 \end{array} \!\right),
\end{equation}
substituted with $p = \widetilde{p} \simeq \pi T$ for small
$|\xi^{\rm i}|$'s. Here,
 the quantity $E_{\mu \mu}= 4
m_{\mu} (m_{\mu} T/2 \pi)^{3/2} \exp(-m_{\mu}/T)$ is the
muon-antimuon energy density, with $m_{\mu}$ as the muon mass,
$m^2_{1,2,3}$ are the squared masses of the three mass
eigenstates, and the transformation between the weak and mass
bases is parameterised with three Euler angles,
\begin{equation}
\label{eq:u}
U \!= \! \left( \!\! \begin{array}{ccc}
                1 & 0 & 0\\
                c_{23} & s_{23} & 0 \\
                -s_{23} & c_{23} & 0 \end{array} \right)\!\!
    \left( \! \! \begin{array}{ccc}
                c_{13} & 0 & s_{13} \\
                0 & 1 & 0 \\
                -s_{13} & 0 & c_{13} \end{array} \right)\!\!
    \left(\! \! \begin{array}{ccc}
                c_{12} & s_{12} & 0 \\
                -s_{12} & c_{12} & 0 \\
                0 & 0 & 1 \end{array} \right),
\end{equation}
where $c_{ij}= \cos \theta_{ij}$, and $s_{ij} = \sin \theta_{ij}$
for $ij=12,\ 23,\ 13$.  Under this scheme, we identify the
atmospheric and solar neutrino oscillation parameters as
\begin{eqnarray}
\Delta m^2_{\rm atm} \equiv m^2_3 - m^2_2, & \qquad & \theta_{\rm
atm } \equiv \theta_{23}, \nonumber \\ \Delta m^2_{\rm sun} \equiv m^2_2 -
m^2_1, & \qquad & \theta_{\rm sun} \equiv \theta_{12},
\end{eqnarray}
and the third angle $\theta_{13}$ is subject to the constraint
$\tan^2 \theta_{13} \alt 0.065$ from a combined analysis of the
solar, atmospheric and CHOOZ data \cite{bib:bestfits}.

Hamiltonian (\ref{eq:3ham}) is simple to decipher, thanks to the
inherent mass hierarchy $\Delta m^2_{\rm atm} \gg \Delta m^2_{\rm
sun}$ (where $\Delta m^2_{\rm atm} \simeq 3 \times 10^{-3}\ {\rm
eV}^2$, and $\Delta m^2_{\rm sun} \alt 10^{-4} \ {\rm eV}^2$), and
the fact that each flavour receives from the background medium a
different contribution to their effective masses.  The evolution
of the three-flavour system follows essentially the dynamics of
three separate and effectively two-flavour subsystems, and the
present parameterisation of the transformation matrix $U$ turns
out to be very convenient for their description
\cite{bib:threeflav}. If all three mixing angles are nonzero, we
have exactly three potentially equilibrating transitions:

(i) The first occurs at $T\simeq 12 \ {\rm MeV}$, when $\Delta
m^2_{\rm atm}/2 \widetilde{p} \simeq (8 \sqrt{2} G_F
\widetilde{p}/3 m^2_W) E_{\mu \mu}$, and $\nu_{\mu}
\leftrightarrow \nu_{\tau}$ oscillations cease to be
matter-suppressed. In a collisionless environment, the subsystem
would undergo an MSW-like transformation which turns $\nu_{\mu}$
and $\nu_{\tau}$ into the states $\nu_x$ and $\nu_y$,
\begin{equation}
 \nu_{\mu} \longrightarrow \nu_x \equiv \frac{1}{\sqrt{2}}
 (\nu_{\mu} - \nu_{\tau}),\quad
\nu_{\tau} \longrightarrow \nu_y \equiv
\frac{1}{\sqrt{2}}(\nu_{\mu} + \nu_{\tau}),
\end{equation}
assuming $\theta_{\rm atm} = 45^{\circ}$.  Collisions play the
role of breaking these states into an incoherent 1:1 mixture of
$\nu_{\mu}$ and $\nu_{\tau}$.

(ii)  The second transition involves the states
\begin{equation}
\nu_e \longrightarrow  c_{13} \nu_e -
s_{13} \nu_y,\qquad
\nu_y  \longrightarrow  s_{13} \nu_e + c_{13} \nu_y,
\end{equation}
and can only be realised for a nonzero $\theta_{13}$.  This
happens when $(\Delta m^2_{\rm atm}+ c^2_{12} \Delta m^2_{\rm
sun})/2 \widetilde{p} \simeq (8 \sqrt{2} G_F \widetilde{p}/3
m^2_W) (E_{ee} + E_{\mu \mu}/2)$, at $T \simeq 5.2\ {\rm MeV}$.
(The factor $1/2$ accompanying $E_{\mu \mu}$ comes from the fact
that the state $\nu_y$ is only ``half-sensitive'' to the
muon-antimuon background.)  In the absence of collisions, Eq.\
(\ref{eq:Iz}) suggests this transition to be quite impotent, even
when $\theta_{13}$ is at its upper limit.   However, noting all
the caveats regarding active-active collisions, a heuristic
approach to collisional damping in the style of Sec.\
\ref{sec:large} can yield for $\tan^2 \theta_{13} \simeq 0.065$ at
$T \simeq 1 \ {\rm MeV}$ an $I_z$ that is less than $1 \%$ of its
original value by Eq.\ (\ref{eq:dampedIz}).  Thus, near
equilibrium between $L_{\nu_e}$ and $L_{\nu_y}$ prior to BBN is
very probable.

(iii) The third transition,
\begin{eqnarray}
c_{13} \nu_e - s_{13} \nu_y
&\longrightarrow & c_{12} (c_{13} \nu_e -
s_{13} \nu_y) - s_{12} \nu_x, \nonumber \\
\nu_x &\longrightarrow&  s_{12}( c_{13} \nu_e - s_{13} \nu_y) +
c_{12} \nu_x,
\end{eqnarray}
was described in detail in Sec.\ \ref{sec:twoflavours} for
$\theta_{13} =0$.  In the case of a sizeable $\theta_{13}$,
equilibrium between $L_{\nu_e}$ and $L_{\nu_y}$ should already be
partially accomplished by this time.  A second equilibrating
transition between $\nu_e$ and $\nu_x$ at this point will  bring
$L_{\nu_e}$, $L_{\nu_{\mu}}$ and $L_{\nu_{\tau}}$ even more in
line.

Note that we have been very specific with the labelling of $\nu_x$
and $\nu_y$.  This actually has an interesting consequence.
Consider the case of a maximal $\theta_{\rm sun}$ and a vanishing
$\theta_{13}$. We may attempt to use a simple ``counting'' method
to establish crudely the  final asymmetries given a set of initial
conditions.  For example, for $L^{\rm i}_{\nu_{e}} = L^{\rm
i}_{\nu_{\tau}} = 0$, and $L^{\rm i}_{\nu_{\mu}} = 0.1$, the first
transition distributes the asymmetry in $\nu_{\mu}$ equally
amongst $\nu_{\mu}$ and $\nu_{\tau}$ such that $L_{\nu_{\mu}}
\simeq L_{\nu_{\tau}} \simeq 0.05$. The second transition is not
present.  At the third transition, the asymmetry carried by
$\nu_x= (\nu_{\mu} - \nu_{\tau})/\sqrt{2}$ is shared evenly with
$\nu_e$, leading to $L_{\nu_e} \simeq L_{\nu_x} \simeq 0.025$.
However, the
 decoupled state $\nu_y=(\nu_{\mu}
+\nu_{\tau})/\sqrt{2}$ still has an asymmetry of $0.05$. Thus the
real $L_{\nu_{\mu}}$ and $L_{\nu_{\tau}}$  should be $\simeq
0.0375$. This simple counting exercise serves to illustrate an
interesting point: without a finite $\theta_{13}$, equilibrium
between $L_{\nu_e}$ and the asymmetries of $\nu_{\mu}$ and
$\nu_{\tau}$ cannot be quite exact.

\section{Conclusion}

We have given an analytical treatment to the neutrino asymmetry
equilibration scenarios considered in the numerical studies of
DHPPRS \cite{bib:dhpprs}. In a two-flavour study of the solar
neutrino oscillation parameters, we demonstrated that the
equilibration mechanism is based upon a collective adiabatic
MSW-like transformation between $\nu_e$ and $\nu_x$ (and between
$\bar{\nu}_e$ and $\bar{\nu}_x$), where $\nu_x$ may be
$\nu_{\mu}$, $\nu_{\tau}$, or a linear combination thereof, for
synchronised $\nu_e \leftrightarrow \nu_x$ and $\bar{\nu}_e
\leftrightarrow \bar{\nu}_x$ oscillations. The transition
temperature is determined, as usual, by $\Delta m^2_{\rm sun}$ and
by a characteristic momentum $\widetilde{p} \simeq \pi T$ for
small initial asymmetries (c.f. $\langle p \rangle \simeq 3.15 \
T$ for a Fermi-Dirac distribution with zero chemical potential).
Thus from the sizes of the various possible $\Delta m^2_{\rm
sun}$'s alone, transition prior to BBN can only be achieved in the
case of the LMA solution.

It turns out that the characteristic momentum $\widetilde{p}$ does
in fact grow with the initial chemical potentials of the ensemble,
and extremely large $|\xi^{\rm i}|$'s can in principle delay the
equilibrating transition. However, the dependence is weak. For
$|\xi^{\rm i}|$ not exceeding $\sim\! 4$, the shift in the
transition temperature is virtually unobservable.

A second concern for the LMA solution is the extent of the
$L_{\nu_e}$/$L_{\nu_x}$ equilibrium .  We showed in the present
work that, when the mixing parameter is chosen to be maximal (as
was done in DHPPRS), complete equilibration of these asymmetries
can always be achieved irrespective of collisional effects, and
for most initial conditions.  (The latter can in principle alter
the adiabaticity of the transition.)  For large but not maximal
mixing, however, the extent of the equilibrium will depend on how
one handles collisions on active-active neutrino oscillations, and
rigorous work on this front is still wanting.  Nonetheless,
heuristic considerations of collisional damping suggest that full
equilibrium before  BBN is most likely.

 For other solutions of the solar neutrino problem, we
 demonstrated in a semi-quantitative three-flavour analysis that
 a partial equilibration of $L_{\nu_e}$ and $L_{\nu_y}$, where
$\nu_y$ is some other linear combination of $\nu_{\mu}$ and
$\nu_{\tau}$, due to a sizeable $\theta_{13}$ is possible in
principle. Again, the extent of this equilibrium hinges on the
treatment of scattering processes with the background, and is
therefore subject to the usual caveats regarding active-active
collisions. Investigations on this topic are currently underway.

(Note: See also Ref.\ \cite{bib:collsynch} for a similar
explanation of the DHPPRS results.)

\acknowledgments{This work was supported by the U. S. Department
of Energy under grant DE-FG02-84ER40163.  The author thanks
K.~N.~Abazajian, N.~F.~Bell, C.~N.~Leung, and R.~R.~Volkas for
discussions and comments on the manuscript, H.~Minakata for
bringing Ref.\ \cite{bib:synch} to her attention, and SISSA, in
particular S.~T.~Petcov, for their hospitality during a recent
visit when the author's interest in the problem arose.}

\end{document}